\documentclass[twocolumn,pra,superscriptaddress,showpacs]{revtex4}
\usepackage[ansinew]{inputenc}
\usepackage[dvips]{graphicx}
\usepackage{amsmath}
\usepackage{amsthm}
\usepackage{layout}
\usepackage{float}
\usepackage{subfigure}
\usepackage{amsfonts}
\usepackage{amssymb}

\begin{document}

\title{Entanglement between Collective Operators in the Linear Harmonic Chain}
\begin{abstract}
We investigate entanglement between collective operators of two blocks of
oscillators in an infinite linear harmonic chain. These operators are defined
as averages over local operators (individual oscillators) in the blocks. On
the one hand, this approach of "physical blocks" meets realistic experimental
conditions, where measurement apparatuses do not interact with single
oscillators but rather with a whole bunch of them, i.e., where in contrast to
usually studied "mathematical blocks" not every possible measurement is
allowed. On the other, this formalism naturally allows the generalization to
blocks which may consist of several non-contiguous regions. We quantify
entanglement between the collective operators by a measure based on the
Peres-Horodecki criterion and show how it can be extracted and transferred to
two qubits. Entanglement between two blocks is found even in the case where
none of the oscillators from one block is entangled with an oscillator from
the other, showing genuine bipartite entanglement between collective
operators. Allowing the blocks to consist of a periodic sequence of subblocks,
we verify that entanglement scales at most with the total boundary region. We
also apply the approach of collective operators to scalar quantum field theory.
\end{abstract}
\date{\today}

%

\author{Johannes Kofler}%

\affiliation
{Institut für Experimentalphysik, Universität Wien, Boltzmanngasse 5, 1090 Wien, Austria}%

\author{Vlatko Vedral}%

\affiliation
{The School of Physics and Astronomy, University of Leeds, Leeds LS2 9JT, United Kingdom}%

\affiliation
{Institut für Experimentalphysik, Universität Wien, Boltzmanngasse 5, 1090 Wien, Austria}%

\author{Myungshik S. Kim}%

\affiliation
{School of Mathematics and Physics, Queen’s University, Belfast BT7 1NN, United Kingdom}%

\author{\v{C}aslav Brukner}%

\affiliation
{Institut für Experimentalphysik, Universität Wien, Boltzmanngasse 5, 1090 Wien, Austria}%

\affiliation
{Institut für Quantenoptik und Quanteninformation, Österreichische Akademie der Wissenschaften,\\ Boltzmanngasse 3, 1090 Wien, Austria}%


\maketitle

\section{Introduction}

Quantum entanglement is a physical phenomenon in which the quantum states of
two or more systems can only be described with reference to each other, even
though the individual systems may be spatially separated. This leads to
correlations between observables of the systems that cannot be understood on
the basis of classical (local realistic) theories \cite{Bell1964}. Its
importance today exceeds the realm of the foundations of quantum physics and
entanglement has become an important physical resource, like energy, that
allows performing communication and computation tasks with efficiency which is
not achievable classically \cite{Niel1998}.

In the near future we will certainly see more and more experiments on
entanglement of increasing complexity. Moving to higher entangled systems or
entangling more systems with each other, will eventually push the realm of
quantum physics well into the macroscopic world. It will be therefore
important to investigate under which conditions entanglement within or between
''macroscopic'' objects, each consisting of a sample containing a large number
of the constituents, can arise.

Recently, it was shown that macroscopic entanglement can arise ''naturally''
between constituents of various complex physical systems. Examples of such
systems are chains of interacting spin systems \cite{Arne2001,Niel1998},
harmonic oscillators \cite{Aude2002,Sera2005} and quantum fields
\cite{Rezn2003}. Entanglement can have an effect on the macroscopic properties
of these systems \cite{Gosh2003,Bruk2005,Wies2005} and can be in principle
extractable from them for quantum information processing
\cite{Rezn2003,Pate2004,deCh2005,Retz2005}.

With the aim of better understanding macroscopical entanglement we will
investigate entanglement between \textit{collective operators} in this paper.
A simple and natural system is the ground state of a linear chain of harmonic
oscillators furnished with harmonic nearest-neighbor interaction. The
mathematical entanglement properties of this system were extensively
investigated in \cite{Aude2002,Sera2005,Pate2005,Bote2004}. Entanglement was
computed in the form of logarithmic negativity for general bisections of the
chain and for contiguous blocks of oscillators that do not comprise the whole
chain. It was shown that the log-negativity typically decreases exponentially
with the separation of the groups and that the larger the groups, the larger
the maximal separation for which the log-negativity is non-zero
\cite{Aude2002}. It also was proven that an area law holds for harmonic
lattice systems, stating that the amount of entanglement between two
complementary regions scales with their boundary \cite{Cram2005}.

In a real experimental situation, however, we are typically not able to
determine the complete mathematical amount of entanglement (as measured, e.g.,
by log-negativity) which is non-zero even if two blocks share only one
arbitrarily weak entangled pair of oscillators. Our measurement apparatuses
normally cannot resolve single oscillators, but rather interact with a whole
bunch of them in one way, potentially even in \textit{non-contiguous regions},
thus measuring certain \textit{global properties}. Here we will study
entanglement between ''physical blocks'' of harmonic oscillators --- existing
only if there is entanglement between the \textit{collective operators}
defined on the entire blocks --- as a function of their size, relative
distance and the coupling strength. Our aim is to quantify (experimentally
accessible) entanglement between global properties of two groups of harmonic
oscillators. Surprisingly, we will see that such collective entanglement can
be demonstrated even in the case where none of the oscillators from one block
is entangled with an oscillator from the other block (i.e., it cannot be
understood as a cumulative effect of entanglement between pairs of
oscillators), which is in agreement with \cite{Aude2002}. This shows the
existence of bipartite entanglement between collective operators.

Because of the area law \cite{Cram2005}\ the amount of entanglement is
relatively small in the first instance. We suggest a way to overcome this
problem by allowing the collective blocks to consist of a \textit{periodic
sequence of subblocks}. Then the total boundary region between them is
increased and we verify that indeed a larger amount of entanglement is found
for periodic blocks, where the entanglement scales at most with the
\textit{total} boundary region. We give an analytical approximation of this
amount of entanglement and motivate how it can in principle be extracted from
the chain \cite{Rezn2003,Pate2004,deCh2005,Retz2005}.

Methodologically, we will quantify the entanglement between collective
operators of two blocks of harmonic oscillators by using a measure for
continuous variable systems based on the Peres-Horodecki criterion
\cite{Pere1996,Horo1997,Simo2000,Kim2002}. The collective operators will be
defined as sums over local operators for all single oscillators belonging to
the block. The infinite harmonic chain is assumed to be in the ground state
and since the blocks do not comprise the whole chain, they are in a mixed state.

\section{Linear Harmonic Chain}

We investigate a linear harmonic chain, where each of the $N$ oscillators is
situated in a harmonic potential with frequency $\omega$ and each oscillator
is coupled with its neighbors by a harmonic potential with the coupling
frequency $\Omega$. The oscillators have mass $m$ and their positions and
momenta are denoted as $\overline{q}_{i}$ and $\overline{p}_{i}$,
respectively. Assuming periodic boundary conditions ($\overline{q}_{N+1}%
\equiv\overline{q}_{1}$), the Hamiltonian thus reads \cite{Schw2003}%
\begin{equation}
H=%
{\displaystyle\sum\limits_{j=1}^{N}}
\left(  \frac{\overline{p}_{j}^{2}}{2\,m}+\frac{m\,\omega^{2}\,\overline
{q}_{j}^{2}}{2}+\frac{m\,\Omega^{2}\,(\overline{q}_{j}-\overline{q}_{j-1}%
)^{2}}{2}\right)  \!.
\end{equation}
We canonically go to dimensionless variables:\ $q_{j}\equiv C\,\overline
{q}_{j}$ and $p_{j}\equiv\overline{p}_{j}/C$, where $C\equiv\sqrt
{m\omega(1+2\,\Omega^{2}/\omega^{2})^{1/2}}$ \cite{Bote2004}. By this means
the Hamiltonian becomes%
\begin{equation}
H=\frac{E_{0}}{2}\,%
{\displaystyle\sum\limits_{j=1}^{N}}
(p_{j}^{2}+q_{j}^{2}-\alpha\,q_{j}\,q_{j+1})\,,
\end{equation}
with the abbreviations $\alpha\equiv2\,\Omega^{2}/(2\,\Omega^{2}+\omega^{2})$
and $E_{0}\equiv\sqrt{2\,\Omega^{2}+\omega^{2}}$. The coupling constant is
restricted to values $0<\alpha<1$, where $\alpha\rightarrow0$ in the weak
coupling limit ($\Omega/\omega\rightarrow0$) and $\alpha\rightarrow1$ in the
strong coupling limit ($\Omega/\omega\rightarrow\infty$).

In the language of second quantization the positions and momenta are converted
into operators ($q_{j}\rightarrow\hat{q}_{j}$, $p_{j}\rightarrow\hat{p}_{j}$)
and are expanded into modes of their annihilation and creation operators,
$\hat{a}$ and $\hat{a}^{\dagger}$, respectively:%
\begin{align}
\hat{q}_{j}  &  =\frac{1}{\sqrt{N}}\,%
{\displaystyle\sum\limits_{k=0}^{N-1}}
\,\frac{1}{\sqrt{2\,\nu(\theta_{k})}}\left[  \hat{a}(\theta_{k})\,\text{e}%
^{\text{i}\,\theta_{k}\,j}+\text{H.c.}\right]  \!,\label{qj}\\
\hat{p}_{j}  &  =\frac{-\text{i}}{\sqrt{N}}\,%
{\displaystyle\sum\limits_{k=0}^{N-1}}
\,\sqrt{\frac{\nu(\theta_{k})}{2}}\left[  \hat{a}(\theta_{k})\,\text{e}%
^{\text{i}\,\theta_{k}\,j}-\text{H.c.}\right]  \!. \label{pj}%
\end{align}
Here $\theta_{k}\equiv2\,\pi\,k/N$ (with $k=0,1,...,N-1$) is the dimensionless
pseudo-momentum and $\nu(\theta_{k})\equiv\sqrt{1-\alpha\cos\theta_{k}}$ is
the dispersion relation. The annihilation and creation operators fulfil the
well known commutation relation $\left[  \hat{a}(\theta_{k}),\hat{a}^{\dagger
}(\theta_{k^{\prime}})\right]  =\delta_{kk^{\prime}}$, since $[\hat{q}%
_{i},\hat{p}_{j}]=\;$i$\,\delta_{ij}$ has to be guaranteed. The ground state
(vacuum), denoted as $\left|  0\right\rangle $, is defined by $\hat{a}%
(\theta_{k})\left|  0\right\rangle =0$ holding for all $\theta_{k}$. The
two-point vacuum correlation functions%
\begin{align}
g_{|i-j|}  &  \equiv\left\langle 0\right|  \hat{q}_{i}\,\hat{q}_{j}\left|
0\right\rangle \equiv\left\langle \!\right.  \hat{q}_{i}\,\hat{q}_{j}\left.
\!\right\rangle ,\label{g}\\
h_{|i-j|}  &  \equiv\left\langle 0\right|  \hat{p}_{i}\,\hat{p}_{j}\left|
0\right\rangle \equiv\left\langle \!\right.  \hat{p}_{i}\,\hat{p}_{j}\left.
\!\right\rangle , \label{h}%
\end{align}
are given by $g_{l}=(2\,N)^{-1}\,%
{\textstyle\sum\nolimits_{k=0}^{N-1}}
\nu^{-1}(\theta_{k})\cos(l\,\theta_{k})$ and $h_{l}=(2\,N)^{-1}\,%
{\textstyle\sum\nolimits_{k=0}^{N-1}}
\nu(\theta_{k})\cos(l\,\theta_{k})$, where $l\equiv|i-j|$. In the limit of an
infinite chain ($N\rightarrow\infty$) --- which we will study below --- and
for $l<N/2$ they can be expressed in terms of the hypergeometric function
$_{2}F_{1}$ \cite{Bote2004}:\ $g_{l}=(z^{l}/2\,\mu)\tbinom{l-1/2}{l}%
\,_{2}F_{1}(1/2,l+1/2,l+1,z^{2})$, $h_{l}=(\mu\,z^{l}/2)\tbinom{l-3/2}%
{l}\,_{2}F_{1}(-1/2,l-1/2,l+1,z^{2})$, where $z\equiv(1-\sqrt{1-\alpha^{2}%
})/\alpha$ and $\mu\equiv1/\sqrt{1+z^{2}}$.

\section{Defining Collective Operators}

In the following, we are interested in entanglement between two ''physical
blocks'' of oscillators, where the blocks are represented by a specific form
of \textit{collective operators} which are normalized sums of individual
operators. By means of such a formalism we seek to fulfil experimental
conditions and constraints, since \textit{finite experimental resolution
implies naturally the measurement of, e.g., the average momentum of a bunch of
oscillators rather than the momentum of only one}. On the other hand, this
formalism can easily take account of blocks that\textit{ consist of
non-contiguous regions}, leading to interesting results which will be shown
below. We want to point out that this convention of the term \textit{block} is
not the same as it is normally used in the previous literature. In contrast to
the latter, for which one allows any possible measurement, our simulation of
realizable experiments already lacks some information due to the averaging.

\begin{figure}[t]
\begin{center}
\includegraphics{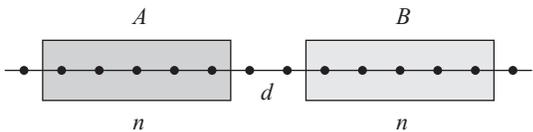}
\end{center}
\par
\vspace{-0.25cm} \caption{Two blocks of a harmonic chain $A$ and $B$. Each
block consists of $n$ oscillators and the blocks are separated by $d$
oscillators.}%
\label{Fig Blocks}%
\end{figure}

Let us now consider two non-overlapping blocks of oscillators, $A$ and $B$,
within the closed harmonic chain in its ground state, where each block
contains $n$ oscillators. The blocks are separated by $d\geq0$ oscillators
(Fig.\ \ref{Fig Blocks}). We assume $n,d\ll N$ and $N\rightarrow\infty$ for
the numerical calculations of the two-point correlation
functions.

By a Fourier transform we map the $n$ oscillators of each block onto $n$
(''orthogonal'') \textit{frequency-dependent} collective operators%
\begin{align}
\hat{Q}_{A}^{(k)}  &  \equiv\dfrac{1}{\sqrt{n}}\;%
{\displaystyle\sum\limits_{j\in A}}
\;\hat{q}_{j}\;\text{e}^{\tfrac{2\,\pi\,\text{i}\,j\,k}{n}},\\
\hat{P}_{A}^{(k)}  &  \equiv\dfrac{1}{\sqrt{n}}\;%
{\displaystyle\sum\limits_{j\in A}}
\;\hat{p}_{j}\;\text{e}^{-\tfrac{2\,\pi\,\text{i}\,j\,k}{n}},
\end{align}
with the frequencies $k=0,...,n-1$, and analogously for block $B$. The
commutator of the collective position and momentum operators is%
\begin{equation}
\lbrack\hat{Q}_{A}^{(k)},\hat{P}_{A}^{(k^{\prime})}]=\text{i}\,\delta
_{kk^{\prime}}\,. \label{commutator QP}%
\end{equation}
This means that collective operators for different frequencies $k\neq
k^{\prime}$ commute. For different blocks the commutator vanishes:\ $[\hat
{Q}_{A}^{(k)},\hat{P}_{B}^{(k^{\prime})}]=0$.

If the individual positions and momenta of all oscillators are written into a
vector%
\begin{equation}
\hat{\mathbf{x}}\equiv(\hat{q}_{1},\hat{p}_{1},\hat{q}_{2},\hat{p}%
_{2},...,\hat{q}_{N},\hat{p}_{N})^{\text{T}},
\end{equation}
then there holds the commutation relation%
\begin{equation}
\lbrack\hat{x}_{i},\hat{x}_{j}]=\text{i}\,\Omega_{ij}%
\end{equation}
with $\mathbf{\Omega}$ the $n$-fold direct sum of $2\!\times\!2$ symplectic
matrices:%
\begin{equation}
\mathbf{\Omega}\equiv%
{\displaystyle\bigoplus\limits_{j=1}^{n}}
\left(  \!%
\begin{array}
[c]{cc}%
0 & 1\\
-1 & 0
\end{array}
\!\right)  \!.
\end{equation}
A matrix $\mathbf{S}$ transforms $\hat{\mathbf{x}}$ into a vector of
collective (and uninvolved individual) oscillators:%
\begin{equation}
\hat{\mathbf{X}}\equiv\mathbf{S}\,\hat{\mathbf{x}}=(\{\hat{Q}_{A}^{(k)}%
,\hat{P}_{A}^{(k)}\}_{k},\{\hat{Q}_{B}^{(k)},\hat{P}_{B}^{(k)}\}_{k},\{\hat
{q}_{j},\hat{p}_{j}\}_{j})^{\text{T}}.
\end{equation}
Here $\{\hat{Q}_{A}^{(k)},\hat{P}_{A}^{(k)}\}_{k}=(\hat{Q}_{A}^{(0)},\hat
{P}_{A}^{(0)},...,\hat{Q}_{A}^{(n-1)},\hat{P}_{A}^{(n-1)})$ denotes all
collective oscillators of block $A$ and analogously for block $B$, whereas
$\{\hat{q}_{j},\hat{p}_{j}\}_{j}$ denotes the $2\,(N-2\,n)$ position and
momentum entries of those $N-2\,n$ oscillators which are not part of one of
the two blocks. The matrix $\mathbf{S}$ corresponds to a Gaussian operation
\cite{Eise2003}. It has determinant det$\,\mathbf{S}=1$ and preserves the
symplectic structure%
\begin{equation}
\mathbf{\Omega}=\mathbf{S}^{\text{T}}\,\mathbf{\Omega}\,\mathbf{S}\,,
\label{sympl struct}%
\end{equation}
and hence%
\begin{equation}
\lbrack\hat{X}_{i},\hat{X}_{j}]=\text{i}\,\Omega_{ij} \label{comm sympl}%
\end{equation}
for all $i,j$, in particular verifying (\ref{commutator QP}). This means that
the Gaussianness of the ground state of the harmonic chain (i.e., the fact
that the state is completely characterized by its first and second moments,
see below) was preserved by the (Fourier) transformation to the
frequency-dependent collective operators.

\section{Quantifying Entanglement between Collective Operators}

In reality, we are typically not capable of single particle resolution
measurements and only of measuring the collective operators with one
frequency, namely $k=0$. Note that in general the correlations of
higher-frequency collective operators, e.g., $\left\langle \!\right.  (\hat
{Q}_{A}^{(k)})^{2}\left.  \!\right\rangle $ or $\left\langle \!\right.
\hat{Q}_{A}^{(k)}\hat{Q}_{B}^{(k)}\left.  \!\right\rangle $ with $k\neq0$, are
not real numbers. Therefore, as a natural choice, we denote as the collective
operators%
\begin{align}
\hat{Q}_{A}  &  \equiv\hat{Q}_{A}^{(0)}=\dfrac{1}{\sqrt{n}}\,%
{\displaystyle\sum\limits_{j\in A}}
\,\hat{q}_{j}\,,\label{def Q}\\
\hat{P}_{A}  &  \equiv\hat{P}_{A}^{(0)}=\dfrac{1}{\sqrt{n}}\,%
{\displaystyle\sum\limits_{j\in A}}
\,\hat{p}_{j}\,, \label{def P}%
\end{align}
and analogously for block $B$. It seems to be a very natural situation that
the experimenter only has access to these collective properties and we are
interested in the amount of (physical) entanglement one can extract from the
system if only the collective observables $\hat{Q}_{A,B}$ and $\hat{P}_{A,B}$
are measured.

Reference \cite{Simo2000} derives a separability criterion which is based on
the Peres-Horodecki criterion \cite{Pere1996,Horo1997} and the fact that ---
in the continuous variables case --- the partial transposition allows a
geometric interpretation as mirror reflection in phase space. Following
largely the notation in the original paper, we introduce the vector%
\begin{equation}
\hat{\mathbf{\xi}}\equiv(\hat{Q}_{A},\hat{P}_{A},\hat{Q}_{B},\hat{P}_{B})
\end{equation}
of collective operators. The commutation relations have the compact form
$[\hat{\xi}_{\alpha},\hat{\xi}_{\beta}]=\;$i$\,K_{\alpha\beta}$ with
$\mathbf{K}\equiv%
{\textstyle\bigoplus\nolimits_{j=1}^{2}}
\!\left(
\genfrac{}{}{0pt}{1}{0}{-1}%
\genfrac{}{}{0pt}{1}{1}{0}%
\right)  $. The separability criterion bases on the covariance matrix (of
first and second moments)%
\begin{equation}
V_{\alpha\beta}\equiv\dfrac{1}{2}\left\langle \!\right.  \Delta\hat{\xi
}_{\alpha}\Delta\hat{\xi}_{\beta}+\Delta\hat{\xi}_{\beta}\Delta\hat{\xi
}_{\alpha}\left.  \!\right\rangle ,
\end{equation}
where $\Delta\hat{\xi}_{\alpha}\equiv\hat{\xi}_{\alpha}-\left\langle
\!\right.  \hat{\xi}_{\alpha}\left.  \!\right\rangle $ with $\left\langle
\!\right.  \hat{\xi}_{\alpha}\left.  \!\right\rangle =0$ in our case (state
around the origin of phase space).

The covariance matrix $\mathbf{V}$ is real (which would not be the case for
higher-frequency collective operators) and symmetric:\ $\left\langle
\!\right.  \hat{Q}_{A}\hat{Q}_{B}\left.  \!\right\rangle =\left\langle
\!\right.  \hat{Q}_{B}\hat{Q}_{A}\left.  \!\right\rangle $ and $\left\langle
\!\right.  \hat{P}_{A}\hat{P}_{B}\left.  \!\right\rangle =\left\langle
\!\right.  \hat{P}_{B}\hat{P}_{A}\left.  \!\right\rangle $, coming from the
fact that the two-point correlation functions (\ref{g}) and (\ref{h}) only
depend on the absolute value of the position index difference. On the other
hand, using (\ref{qj}) and (\ref{pj}), we verify that $\left\langle \!\right.
\hat{q}_{i}\,\hat{p}_{j}\left.  \!\right\rangle =\;$i$\,(2\,N)^{-1}\,%
{\textstyle\sum\nolimits_{k=0}^{N-1}}
\exp[$i$\,\theta_{k}(i-j)]$ and $\left\langle \!\right.  \hat{p}_{j}\,\hat
{q}_{i}\left.  \!\right\rangle =-$i$\,(2\,N)^{-1}\,%
{\textstyle\sum\nolimits_{k=0}^{N-1}}
\exp[$i$\,\theta_{k}(j-i)]$. For $i\neq j$ both summations vanish ($\theta
_{k}\equiv2\,\pi\,k/N$ and $i,j$ integer) and for $i=j$ they are the same but
with opposite sign. Thus, in all cases $\left\langle \!\right.  \hat{q}%
_{i}\,\hat{p}_{j}\left.  \!\right\rangle =-\left\langle \!\right.  \hat{p}%
_{j}\,\hat{q}_{i}\left.  \!\right\rangle $. These symmetries also hold for the
collective operators and hence we obtain%
\begin{equation}
\mathbf{V}=\left(  \!%
\begin{array}
[c]{cccc}%
G & 0 & G_{AB} & 0\\
0 & H & 0 & H_{AB}\\
G_{AB} & 0 & G & 0\\
0 & H_{AB} & 0 & H
\end{array}
\!\right)  \!. \label{V}%
\end{equation}
The matrix elements are%
\begin{align}
G  &  \equiv\left\langle \!\right.  \hat{Q}_{A}^{2}\left.  \!\right\rangle
=\left\langle \!\right.  \hat{Q}_{B}^{2}\left.  \!\right\rangle =\frac{1}{n}\,%
{\displaystyle\sum\limits_{j\in A}}
\,%
{\displaystyle\sum\limits_{i\in A}}
\,g_{|j-i|}\,,\\
H  &  \equiv\left\langle \!\right.  \hat{P}_{A}^{2}\left.  \!\right\rangle
=\left\langle \!\right.  \hat{P}_{B}^{2}\left.  \!\right\rangle =\frac{1}{n}\,%
{\displaystyle\sum\limits_{j\in A}}
\,%
{\displaystyle\sum\limits_{i\in A}}
\,h_{|j-i|}\,,\\
G_{AB}  &  \equiv\left\langle \!\right.  \hat{Q}_{A}\hat{Q}_{B}\left.
\!\right\rangle =\frac{1}{n}\,%
{\displaystyle\sum\limits_{j\in A}}
\,%
{\displaystyle\sum\limits_{i\in B}}
\,g_{|j-i|}\,,\\
H_{AB}  &  \equiv\left\langle \!\right.  \hat{P}_{A}\hat{P}_{B}\left.
\!\right\rangle =\frac{1}{n}\,%
{\displaystyle\sum\limits_{j\in A}}
\,%
{\displaystyle\sum\limits_{i\in B}}
\,h_{|j-i|}\,.
\end{align}
To quantify entanglement between two collective blocks we use the degree of
entanglement $\varepsilon$, given by the absolute sum of the negative
eigenvalues of the partially transposed density operator:\ $\varepsilon
\equiv\;$Tr$|\rho^{\text{T}_{B}}|-1$, i.e., by measuring how much the mirror
reflected state fails to be positive definite. This measure (negativity) is
based on the Peres-Horodecki criterion \cite{Pere1996,Horo1997} and was shown
to be an entanglement monotone \cite{Lee2000,Vida2002}. For covariance
matrices of the form (\ref{V}) it reads \cite{Kim2002}%
\begin{equation}
\varepsilon=\max\left(  0,\frac{(\delta_{1}\delta_{2})_{0}}{\delta_{1}%
\delta_{2}}-1\right)  \!, \label{epsilon}%
\end{equation}
where $\delta_{1}\equiv G-|G_{AB}|$ and $\delta_{2}\equiv H-|H_{AB}|$. In
general, the numerator is defined by the square of the Heisenberg uncertainty
relation%
\begin{equation}
(\delta_{1}\delta_{2})_{0}\equiv\left(  \frac{1}{2}\,|\!\left\langle
\!\right.  [\hat{Q}_{A,B},\hat{P}_{A,B}]\left.  \!\right\rangle \!|\right)
^{\!2},
\end{equation}
with $(\delta_{1}\delta_{2})_{0}=1/4$ due to (\ref{commutator QP}). We note
that $\varepsilon$ is a \textit{degree} of entanglement (in the sense of
necessity and sufficiency) only for Gaussian states which are completely
characterized by their first and second moments, as for example the ground
state of the harmonic chain we are studying. However, we left out the
higher-frequency collective operators (and all the oscillators which are not
part of the blocks) and therefore, the entanglement $\varepsilon$ has to be
understood as the Gaussian part of the amount of entanglement which exists
between (and can be extracted from) the two blocks when only the collective
properties $\hat{Q}_{A,B}$ and $\hat{P}_{A,B}$, as defined in (\ref{def Q})
and (\ref{def P}), are accessible.

There also exists an entanglement witness in form of a separability criterion
based on variances, where $\Delta\equiv\left\langle \!\right.  (\hat{Q}%
_{A}-\hat{Q}_{B})^{2}\left.  \!\right\rangle +\left\langle \!\right.  (\hat
{P}_{A}+\hat{P}_{B})^{2}\left.  \!\right\rangle =2\,(G-G_{AB}+H+H_{AB})<2$ is
a sufficient condition for the state to be entangled \cite{Duan2000}. We note
that the above negativity measure (\ref{epsilon}) is ''stronger'' than this
witness in the whole parameter space ($\alpha,n$). In particular, there are
cases where $\varepsilon>0$ although $\Delta\geq2$. This is in agreement with
the finding that the variance criterion is weaker than a generalized
negativity criterion \cite{Shch2005}.

We further note that the amount of entanglement (\ref{epsilon}) is invariant
under a change of potential redefinitions of the collective operators, e.g.,
$\hat{Q}_{A}\equiv\,%
{\textstyle\sum\nolimits_{j\in A}}
\,\hat{q}_{j}$ or $\hat{Q}_{A}\equiv(1/n)\,%
{\textstyle\sum\nolimits_{j\in A}}
\,\hat{q}_{j}$, as then the modified scaling in the correlations ($G$,
$G_{AB}$, $H$, and $H_{AB}$) is exactly compensated by the modified scaling of
the Heisenberg uncertainty in the numerator.

\begin{figure}[t]
\begin{center}
\includegraphics{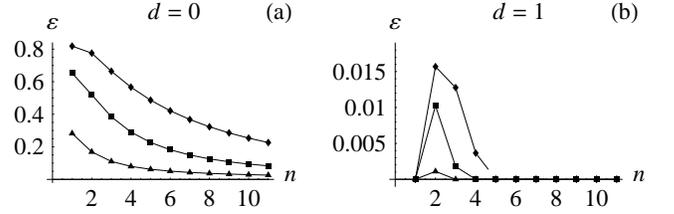}
\end{center}
\par
\vspace{-0.25cm}\caption{Degree of collective entanglement $\varepsilon$ for
two blocks of oscillators as a function of their size $n$. (a) The blocks are
neighboring ($d=0$) and entanglement exists for all $n$ and coupling strengths
$\alpha$. Plotted are $\alpha=0.99$ (diamonds), $\alpha=0.9$ (squares) and
$\alpha=0.5$ (triangles). (b) The same for two blocks which are separated by
one oscillator ($d=1$). The two blocks are unentangled for $n=1$ but can be
entangled, if one increases the block size ($n>1$), although non of the
individual pairs between the blocks is entangled.}%
\label{Fig Degr}%
\end{figure}

Figure \ref{Fig Degr} shows the results for $d=0$ and $d=1$. In the first case
--- if the blocks are neighboring --- there exists entanglement for all
possible coupling strengths $\alpha$ and block sizes $n$. In the latter case
--- if there is one oscillator between the blocks --- due to the strongly
decaying correlation functions $g$ an $h$ there is no entanglement between two
single oscillators ($n=1$), but entanglement for larger blocks (up to $n=4$,
depending on $\alpha$). The statement that entanglement can emerge by going to
larger blocks was also found in \cite{Aude2002}. But there the blocks were
abstract objects, containing all the information of their constituents. In the
case of collective operators, however, increasing the block size (averaging
over more oscillators) is also connected with a loss of information. In spite
of this loss and the mixedness of the state, two blocks can be entangled,
although non of the individual pairs between the blocks is entangled ---
indicating true bipartite entanglement between collective operators. For
$d\geq2$, however, no entanglement can be found anymore.

\begin{figure}[t]
\begin{center}
\includegraphics{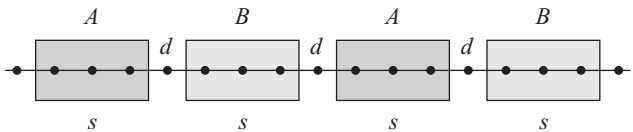}
\end{center}
\par
\vspace{-0.25cm} \caption{Two periodic blocks of a harmonic chain $A$ and $B$.
Each block can consist of $m$ subblocks with $s$ oscillators each, separated
by $d$ oscillators. In the picture $d=1$, $m=2$, $s=3$ and the number of
oscillators per block is $n=ms=6$.}%
\label{Fig Blocksgen}%
\end{figure}

These results are in agreement with the general statement that entanglement
between a region and its complement scales with the size of the boundary
\cite{Cram2005}. In the present case of two blocks in a one-dimensional chain
(Fig.\ \ref{Fig Blocks}) the boundary is constant and as the blocks are made
larger, the entanglement decreases since it is distributed over more and more
oscillators. We therefore propose to increase the number of boundaries by
considering two non-overlapping blocks, where we allow a \textit{periodic
continuation} of the situation above, i.e.\ a sequence of $m\geq1$ subblocks,
separated by $d$ oscillators and each consisting of $s\geq1$ oscillators,
where $ms=n$ (Fig.\ \ref{Fig Blocksgen}).

\begin{figure}[t]
\begin{center}
\includegraphics{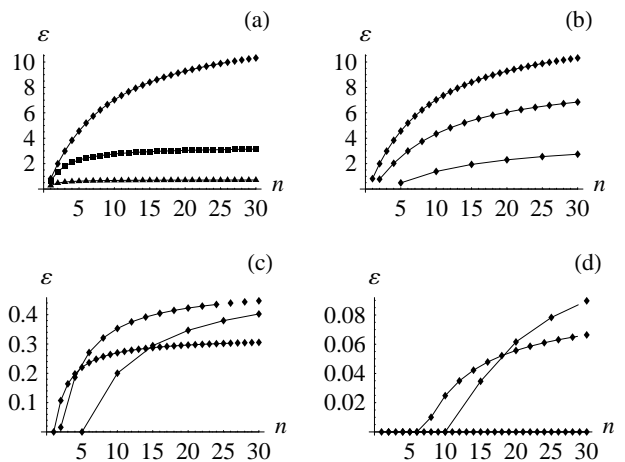}
\end{center}
\par
\vspace{-0.25cm}\caption{Degree of collective entanglement $\varepsilon$ for
two periodic blocks of oscillators as a function of their total size $n$. (a)
Neighboring one-particle subblocks ($d=0$, $s=1$). Entanglement monotonically
increases with $n$ and becomes larger as the coupling strength $\alpha$
increases. Plotted are $\alpha=0.99$ (diamonds), $\alpha=0.9$ (squares) and
$\alpha=0.5$ (triangles). (b) The coupling is fixed to $\alpha=0.99$ for this
and the subsequent figures. There is no separation, $d=0$. Plotted are the
cases $s=1,2,5$. For fixed $n$ the entanglement is more or less proportional
to the number of boundaries, i.e., inversely proportional to the subblock size
$s$. (c) and (d) correspond to the cases $d=1$ and $d=2$, respectively. The
dependence on the size of the subblocks is more complicated as there is a
trade-off between having a large number of boundaries (i.e.\ small $s$) and
the fact that one should have large subblocks as individual separated
oscillators are not entangled.}%
\label{Fig Degrgen}%
\end{figure}

The degree of entanglement between two periodic blocks of non-separated
($d=0$) one-particle subblocks ($s=1$) is larger for stronger coupling
constant $\alpha$ and grows with the overall number of oscillators $n$
(Fig.\ \ref{Fig Degrgen}a). For given $\alpha$ and $n$ and no separation
between the subblocks ($d=0$) the entanglement is larger for the case of small
subblocks, as then there are many of them, causing a large total boundary
(Fig.\ \ref{Fig Degrgen}b). Entanglement can be even found for larger
separation ($d=1,2$) with a more complicated dependence on the size $s$ of the
subblocks. There is a trade-off between having a large number of boundaries
and the fact that one should have large subblocks as individual separated
oscillators are not entangled (Fig.\ \ref{Fig Degrgen}c,d). For $d\geq3$ no
entanglement can be found anymore. (In a realistic experimental situation,
where the separation $d$ is not sharply defined, e.g., where there are
weighted contributions for $d=0,1,...,d_{\text{max}}$, entanglement can
persist even for $d_{\text{max}}\geq3$, depending on the weighting
factors.)

For the sake of completeness we give a rough approximation of the entanglement
between two periodic blocks. Let us assume that the subblocks are directly
neighbored, $d=0$. Furthermore, we consider couplings $\alpha$ such that we
may neglect higher than next neighbor correlations ($\alpha\lesssim0.5$),
i.e., we only take into account $g_{0}$, $g_{1}$, $h_{0}$ and $h_{1}$. The
correlations read%
\begin{align}
G  &  =\dfrac{1}{n}\,%
{\displaystyle\sum\limits_{j\in A}}
\,%
{\displaystyle\sum\limits_{i\in A}}
\,g_{|j-i|}\approx g_{0}+\dfrac{2\,m\,(s-1)}{n}\,g_{1}\,,\\
G_{AB}  &  =\dfrac{1}{n}\,%
{\displaystyle\sum\limits_{j\in A}}
\,%
{\displaystyle\sum\limits_{i\in B}}
\,g_{|j-i|}\approx\dfrac{1}{n}\,(2\,m-1)\,g_{1}\,,
\end{align}
and analogously for $H$ and $H_{AB}$. The first equation reflects that there
are $n$ self-correlations and $m\,(s-1)$ nearest neighbor pairs (which are
counted twice) within one block, i.e., $s-1$ pairs per subblock. The second
equation represents the fact that there are $2\,m-1$ boundaries where blocks
$A$ and $B$ meet. Using $s=n/m$, the entanglement (\ref{epsilon}) becomes
(note that $g_{1}>0$ and $h_{1}<0$)%
\begin{equation}
\varepsilon\approx\dfrac{1}{4\,[g_{0}+(2-\tfrac{4\,m-1}{n})\,g_{1}%
]\,[h_{0}+(2-\tfrac{1}{n})\,h_{1}]}-1\,.
\end{equation}
For given $n$ this approximation obviously increases with the total number of
boundaries, $m$. It can be considered as an estimate for a situation like in
Fig.\ \ref{Fig Degrgen}b, if a smaller coupling is used such that the neglect
of higher correlations becomes justified.

We close this section by annotating that the entanglement (\ref{epsilon})
between collective blocks of oscillators --- being the Gaussian part --- can
in principle (for sufficient control of the block separation $d$) be
transferred to two remote qubits via a Jaynes--Cummings type interaction
\cite{Rezn2003,Pate2004,Retz2005}. For the interaction with periodic blocks
''gratings'' have to be employed in the experimental setup. The interaction
Hamiltonian is of the form%
\begin{align}
\hat{H}_{\text{int}}  &  \sim(\text{e}^{-\text{\thinspace i\thinspace}%
\omega_{1}\,t}\,\hat{\sigma}_{1}^{+}+\text{e}^{+\text{\thinspace i\thinspace
}\omega_{1}\,t}\,\,\hat{\sigma}_{1}^{+})\,\hat{Q}_{A}\nonumber\\
&  \quad\quad+(\text{e}^{-\text{\thinspace i\thinspace}\omega_{2}\,t}%
\,\hat{\sigma}_{2}^{+}+\text{e}^{+\text{\thinspace i\thinspace}\omega_{2}%
\,t}\,\hat{\sigma}_{2}^{+})\,\hat{Q}_{B}\,,
\end{align}
where $\omega_{i}$ is the Rabi frequency and $\hat{\sigma}_{i}^{+}%
=(\hat{\sigma}_{i}^{-})^{\dagger}=\left|  e\right\rangle \!_{i}\,_{i}%
\!\left\langle g\right|  $ is the bosonic operator (with $\left|
g\right\rangle \!_{i}$ and $\left|  e\right\rangle \!_{i}$ the ground and the
excited state) of the $i$-th qubit ($i=1,2$).

\section{Collective Operators for Scalar Quantum Fields}

The continuum limit of the linear harmonic chain is the (1+1)-dimensional
Klein--Gordon field $\phi(x,t)$ with the canonical momentum field
$\pi(x,t)=\dot{\phi}(x,t)$. It satisfies the Klein--Gordon equation (in
natural units $\hbar=c=1$) with mass $m$%
\begin{equation}
\ddot{\phi}-\nabla^{2}\phi+m^{2}\,\phi=0\,.
\end{equation}
With the canonical quantization procedure $\phi$ and $\pi$ become operators
satisfying the non-trivial commutation relation $[\hat{\phi}(x,t),\hat{\pi
}(x^{\prime},t)]=\;$i$\,\delta(x-x^{\prime})$. The field operator can be
expanded into a Fourier integral over elementary plane wave solutions
\cite{Bjo2003}%
\begin{align}
\hat{\phi}(x,t)  &  =%
{\displaystyle\int}
\frac{\text{d}k}{\sqrt{4\,\pi\,\omega_{k}}}\left[  \hat{a}(k)\,\text{e}%
^{\text{i}\,k\,x-\text{i}\,\omega_{k}\,t}+\text{H.c.}\right]  \!,\label{field}%
\\
\hat{\pi}(x,t)  &  =-\text{i}%
{\displaystyle\int}
\frac{\text{d}k\,\omega_{k}}{\sqrt{4\,\pi}}\left[  \hat{a}(k)\,\text{e}%
^{\text{i}\,k\,x-\text{i}\,\omega_{k}\,t}-\text{H.c.}\right]  \!,
\end{align}
where $k$ is the wave number and $\omega_{k}=+\sqrt{k^{2}+m^{2}}$ is the
dispersion relation. The annihilation and creation operators fulfil $\left[
\hat{a}(k),\hat{a}^{\dagger}(k^{\prime})\right]  =\delta(k-k^{\prime})$. We
write the field operator as a sum of two contributions $\hat{\phi}=\hat{\phi
}^{(+)}+\hat{\phi}^{(-)}$, where $\hat{\phi}^{(+)}$ ($\hat{\phi}^{(-)}$) is
the contribution with positive (negative) frequency. Thus, $\hat{\phi}^{(+)}$
corresponds to the term with the annihilation operator in (\ref{field}). The
vacuum correlation function is given by the (equal-time) commutator of the
positive and the negative frequency part:%
\begin{equation}
\left\langle 0\right|  \hat{\phi}(x,t)\,\hat{\phi}(y,t)\left|  0\right\rangle
=[\hat{\phi}^{(+)}(x,t),\hat{\phi}^{(-)}(y,t)]\,.
\end{equation}
It is a peculiarity of the idealization of quantum field theory that for $x=y$
this propagator diverges in the ground state:%
\begin{equation}
\left\langle 0\right|  \hat{\phi}^{2}(x,t)\left|  0\right\rangle
\rightarrow\infty\,. \label{divergence}%
\end{equation}
The same is true for $\left\langle 0\right|  \hat{\pi}^{2}(x,t)\left|
0\right\rangle $ and hence we cannot easily build an entanglement measure like
for the harmonic chain, since the analogs of the two-point correlation
functions $g_{0}$ and $h_{0}$, (\ref{g}) and (\ref{h}), are divergent now.
Automatically, we are motivated to study the more physical situation and
consider extended space-time regions, which means that we should integrate the
field (and conjugate momentum) over some spatial area. We define the
collective operators%
\begin{align}
\hat{\Phi}_{L}(x_{0},t)  &  \equiv\frac{1}{\sqrt{L}}\,%
{\displaystyle\int\nolimits_{-L/2}^{L/2}}
\,\hat{\phi}(x+x_{0},t)\,\text{d}x\,,\label{Phi}\\
\hat{\Pi}_{L}(x_{0},t)  &  \equiv\frac{1}{\sqrt{L}}\,%
{\displaystyle\int\nolimits_{-L/2}^{L/2}}
\,\hat{\pi}(x+x_{0},t)\,\text{d}x\,, \label{Pi}%
\end{align}
Therefore, $\hat{\Phi}_{L}(x_{0},t)$ and $\hat{\Pi}_{L}(x_{0},t)$ are
equal-time operators which are spatially averaged over a length $L$, centered
at position $x_{0}$. The commutator is%
\begin{equation}
\lbrack\hat{\Phi}_{L}(x_{0},t),\hat{\Pi}_{L}(x_{0},t)]=\frac{1}{L}\,%
{\displaystyle\int\nolimits_{-L/2}^{L/2}}
\,%
{\displaystyle\int\nolimits_{-L/2}^{L/2}}
\,\text{i}\,\delta^{(3)\!}(x-y)=\text{i}\,, \label{commutator PhiPi}%
\end{equation}
which is in complete analogy to (\ref{commutator QP}). If $\hat{\Phi}_{L}$ and
$\hat{\Pi}_{L}$ correspond to separated regions without overlap, i.e.,
$|x_{0}-y_{0}|>L$, then of course $[\hat{\Phi}_{L}(x_{0},t),\hat{\Pi}%
_{L}(y_{0},t)]=0$. The spatial integration in (\ref{Phi}) and (\ref{Pi}) can
be carried out analytically:%
\begin{align}
\hat{\Phi}_{L}(x_{0},t)  &  =\frac{1}{\sqrt{\pi\,L}}\,%
{\displaystyle\int\nolimits_{-\infty}^{\infty}}
\,\frac{\text{d}k}{k\,\sqrt{\omega_{k}}}\,\sin\!\left(  \frac{k\,L}{2}\right)
\nonumber\\
&  \quad\times\left[  \hat{a}(k)\,\text{e}^{\text{i}\,k\,x_{0}-\text{i}%
\,\omega_{k}\,t}+\text{H.c.}\right]  \!,\label{field Phi}\\
\hat{\Pi}_{L}(x_{0},t)  &  =\frac{-\text{i}}{\sqrt{\pi\,L}}\,%
{\displaystyle\int\nolimits_{-\infty}^{\infty}}
\,\frac{\text{d}k\,\sqrt{\omega_{k}}}{k}\,\sin\!\left(  \frac{k\,L}{2}\right)
\nonumber\\
&  \quad\times\left[  \hat{a}(k)\,\text{e}^{\text{i}\,k\,x_{0}-\text{i}%
\,\omega_{k}\,t}-\text{H.c.}\right]  \!. \label{field Pi}%
\end{align}
The final step is to calculate the propagators of the field and the conjugate
momentum.\ We find%
\begin{align}
D_{\hat{\Phi},L}(r)  &  \equiv\left\langle 0\right|  \hat{\Phi}_{L}%
(x_{0},t)\,\hat{\Phi}_{L}(y_{0},t)\left|  0\right\rangle \label{DPhi}\\
&  =\frac{1}{\pi\,L}\,%
{\displaystyle\int\nolimits_{-\infty}^{\infty}}
\,\frac{\text{d}k}{k^{2}\,\sqrt{k^{2}+m^{2}}}\,\sin^{2}\!\left(  \frac
{k\,L}{2}\right)  \cos(k\,r)\,,\nonumber\\
D_{\hat{\Pi},L}(r)  &  \equiv\left\langle 0\right|  \hat{\Pi}_{L}%
(x_{0},t)\,\hat{\Pi}_{L}(y_{0},t)\left|  0\right\rangle \label{DPi}\\
&  =\frac{1}{\pi\,L}\,%
{\displaystyle\int\nolimits_{-\infty}^{\infty}}
\,\frac{\text{d}k\,\sqrt{k^{2}+m^{2}}}{k^{2}}\,\sin^{2}\!\left(  \frac
{k\,L}{2}\right)  \cos(k\,r)\,,\nonumber
\end{align}
with $r\equiv|x_{0}-y_{0}|$ the distance between the centers of the two
regions, reflecting the spatial symmetry. Thus $D_{\hat{\Phi},L}(0)$ and
$D_{\hat{\Pi},L}(0)$ are the analogs of $\left\langle \!\right.  \hat{Q}%
_{A,B}^{2}\left.  \!\right\rangle $ and $\left\langle \!\right.  \hat{P}%
_{A,B}^{2}\left.  \!\right\rangle $ (\textit{intra}-block correlations within
the same block), respectively, whereas $D_{\hat{\Phi},L}(r>L)$ and
$D_{\hat{\Pi},L}(r>L)$ correspond to $\left\langle \!\right.  \hat{Q}_{A}%
\hat{Q}_{B}\left.  \!\right\rangle $ and $\left\langle \!\right.  \hat{P}%
_{A}\hat{P}_{B}\left.  \!\right\rangle $ (\textit{inter}-block correlations
between separated blocks).

The expressions (\ref{DPhi}) and (\ref{DPi}) are finite, especially for $r=0$.
Mathematically, the integration over a finite spatial region $L$ corresponds
to a cutoff, which removes the divergence we faced in (\ref{divergence}).
However, the expressions are ill defined for $L\rightarrow0$.

Applying the entanglement measure (\ref{epsilon}) with $G=D_{\hat{\Phi},L}%
(0)$, $H=D_{\hat{\Pi},L}(0)$, $G_{AB}=D_{\hat{\Phi},L}(r)$, and $H_{AB}%
=D_{\hat{\Pi},L}(r)$ does not indicate entanglement for any choice of $L$ and
$r>L$. The same is true for the generalized case of blocks consisting of
periodic subregions of space, showing an inherent difference between the
harmonic chain and its continuum limit. We believe this is due to the fact,
that any spatial integration immediately corresponds to an infinitely large
block in the discrete harmonic chain and that the information loss (compared
to the mathematical indeed existing exponentially small entanglement
\cite{Rezn2003}) due to the collective operators already is too large.
Nonetheless, defining collective operators like in (\ref{Phi}) and (\ref{Pi})
and use of the measure (\ref{epsilon}) may reveal entanglement between
spatially separated regions for other quantum field states, which is the
subject of future research.

\section{Conclusion}

Our results have importance for investigating the conditions under which
entanglement can be detected by measuring collective observables of blocks
consisting of a large number of harmonic oscillators. This has relevance for
schemes of extracting entanglement where the probe particles normally interact
with whole (periodic) groups of oscillators rather than single oscillators.
The results are also relevant for the transition from the quantum to the
classical domain as they suggest that entanglement between collective
operators (global properties) may persist even in the limit of a large number
of particles. It is obvious that our approach of collective observables can be
extended to more dimensions. Furthermore, we demonstrated its potential
application to scalar quantum field theory.

\section*{Acknowledgement}

We thank A.\ Ferreira and M.\ Wie\'{s}niak for stimulating discussions.
J.\ K.\ and \v{C}.\ B.\ acknowledge financial support by the FWF (Austrian
Science Fund), SFB project P06,\ as well as the European Commission under the
Integrated Project Qubit Applications QAP funded by the IST Directorate as
contract number 015846. V.\ V.\ acknowledges support from the Engineering and
Physical Sciences Council, the British Council in Austria and the European Union.

\end{document}